\documentclass[
aps,%
12pt,%
final,%
notitlepage,%
oneside,%
onecolumn,%
nobibnotes,%
nofootinbib,%
superscriptaddress,%
noshowpacs,%
centertags]%
{revtex4}
\usepackage{amssymb}
\usepackage{amsmath}
\usepackage{epsfig}
\usepackage{multirow}
\usepackage[cp866]{inputenc}
\usepackage[russian]{babel}
\newcommand{\be}{\begin{equation}}
\newcommand{\ee}{\end{equation}}
\newcommand{\bea}{\begin{eqnarray}}
\newcommand{\eea}{\end{eqnarray}}
\newcommand{\m}{\mathbf}
\begin{document}
\title{\bf Кулоновская дезинтеграция тяжелых ионов для $^{208}$Pb 2$\times$1.38~ГэВ/c   }
\author{Ф.Карпешин\\ Санкт-Петербургский государственный универсистет\\ }
\begin{abstract}
Проанализирован спектр виртуальных фотонов для ядер Pb-208,
сталкивающихся в большом адроном коллайдере. Вычислены сечения
неупругих кулоновских сnолкновений с испарением одного, двух и
трех нейтронов одним из партнеров. Методом Глаубера выполнены
оценки множественности нейтронов, возникающих в результате
кулоновского разрушения ядер. Предсказанная нейтронная
множественность достигает 30--40 и больше для каждого из
партнеров. Нейтроны сопровождаются испущенными более быстрыми
протонами и другими заряженными частицами, $\emph{в среднем}$
тремя-пятью частицами для каждого партнера. Большинство частиц
кулоновского происхождения характеризуются  малыми рапидити и
могут, таким образом, быть зарегистрированы в ZDC.
\end{abstract}

\maketitle

\section{Физические предпосылки}

\begin{figure}
\centerline{
\epsfxsize=11cm \epsfysize=7cm \epsfbox{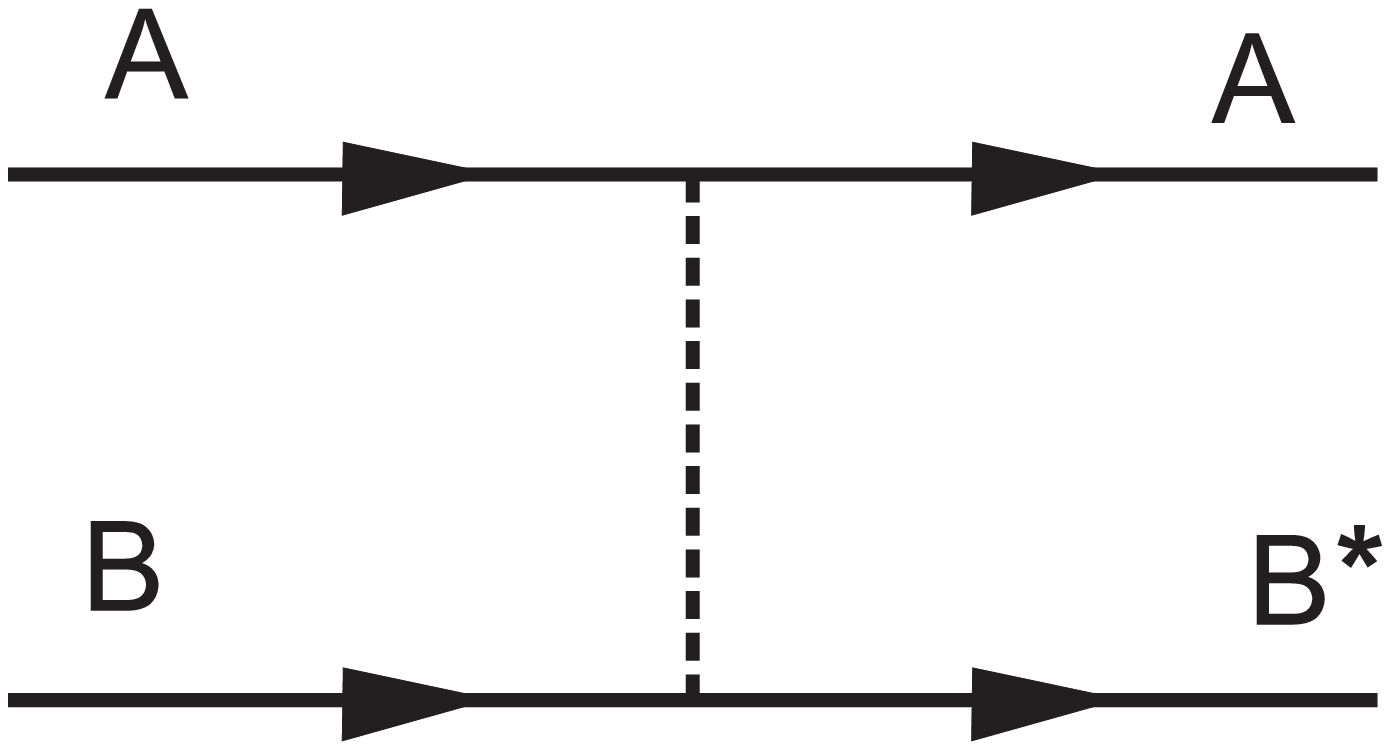}}
\caption{\footnotesize Фейнмановский график процесса столкновения
ионов $A$ и $B$ с возбуждением одного из них ($B$).} \label{f1gr}
\end{figure}

    Как известно, столкновение тяжелых ионов можт сопровождаться их взаимным кулоновским возбуждением, даже если прицельные параметры за пределами взаимодействия ядерных сил. Амплитуда возбуждения ядра-мишени кулоновским полем налетающего иона в первом приближении  дается фейнмановским графиком на рис. \ref{f1gr}. Она может быть вычислена методом эквивалентных фотонов Вейцзэкера-Вильямса. Его квантовоэлектродинамическая формулировка была дана Грибовым \cite{gri}. Основной вклад в сечение происходит от области почти реальных
фотонов в промежуточном состоянии с $q^2 \lesssim 0$,  $ q = (\omega, \m k)$ --- четырехимпульс фотона. Сечение процесса на рис. \ref{f1gr} равно
\be \frac {d\sigma}{d\omega} = \frac {d\sigma_{ph}}{d\omega}
n(\omega) \,, \label{vf1} \ee где $\frac {d\sigma_{ph}}{d\omega}$
--- сечение фотовозбуждения ядра, $n(\omega)$ --- спектр
виртуальных фотонов. В приближении кулоновского центра последний
дается формулой \cite{gri}
\be n(\omega) =
\frac{Z^2\alpha}{\pi\omega} \ln \frac{p^2}{ \omega^2}\,,
\label{vf2} \ee
где $p$ --- импульс налетающего ядра. Он же
определяет  кинематический предел спектра виртуальных фотонов, то
есть максимально возможную энергию возбуждения ядра-мишени. Для
энергий ускорения ионов свинца $^{208}$Pb на БАКе
2$\times$1.38~МэВ/нуклон, в лабораторной системе лоренц-фактор при
энергии БАКа  $E$ = 1.38 ТэВ на нуклон равен $\gamma_L^\prime = E/
m_N$ = 1470, $m_N$ --- нуклонная масса.  В системе покоя одного из
ядер, которое будем называить условно ядром-мишенью, лоренц-фактор
становится $\gamma_L = 2 {\gamma_L^\prime}^2 - 1 = 4.32\times
10^6$. В результате получим величину $p = A m_N \gamma_L =
2.1\times 10^5$ ТэВ.

    Конечные размеры сталкивающихся ядер, однако, важны для определения параметров спектра ВФ. Их учет можно произвести методом Фурье-преобразования запаздывающего потенциала от налетающего иона \cite{J}. Пусть прицельное расстояние налетающего иона   $b$. Тогда можно записать следующее выражение для потока энергии в системе покоя ядра-мишени \cite{J,mp}
\be
\frac{d^3 S}{dk d^2b} =  \frac{Z^2 \alpha w^2}{ (\pi r)^2}
\left[ K_1^2 (w) + \frac1{\gamma_L^2} K_0^2(w)\right]\,.  \label{nl}
\ee
где $w = \frac{kr}{\gamma_L}$.
\begin{figure}[!hbt]
\centerline{
\epsfxsize=8cm \epsfysize=6cm \epsfbox{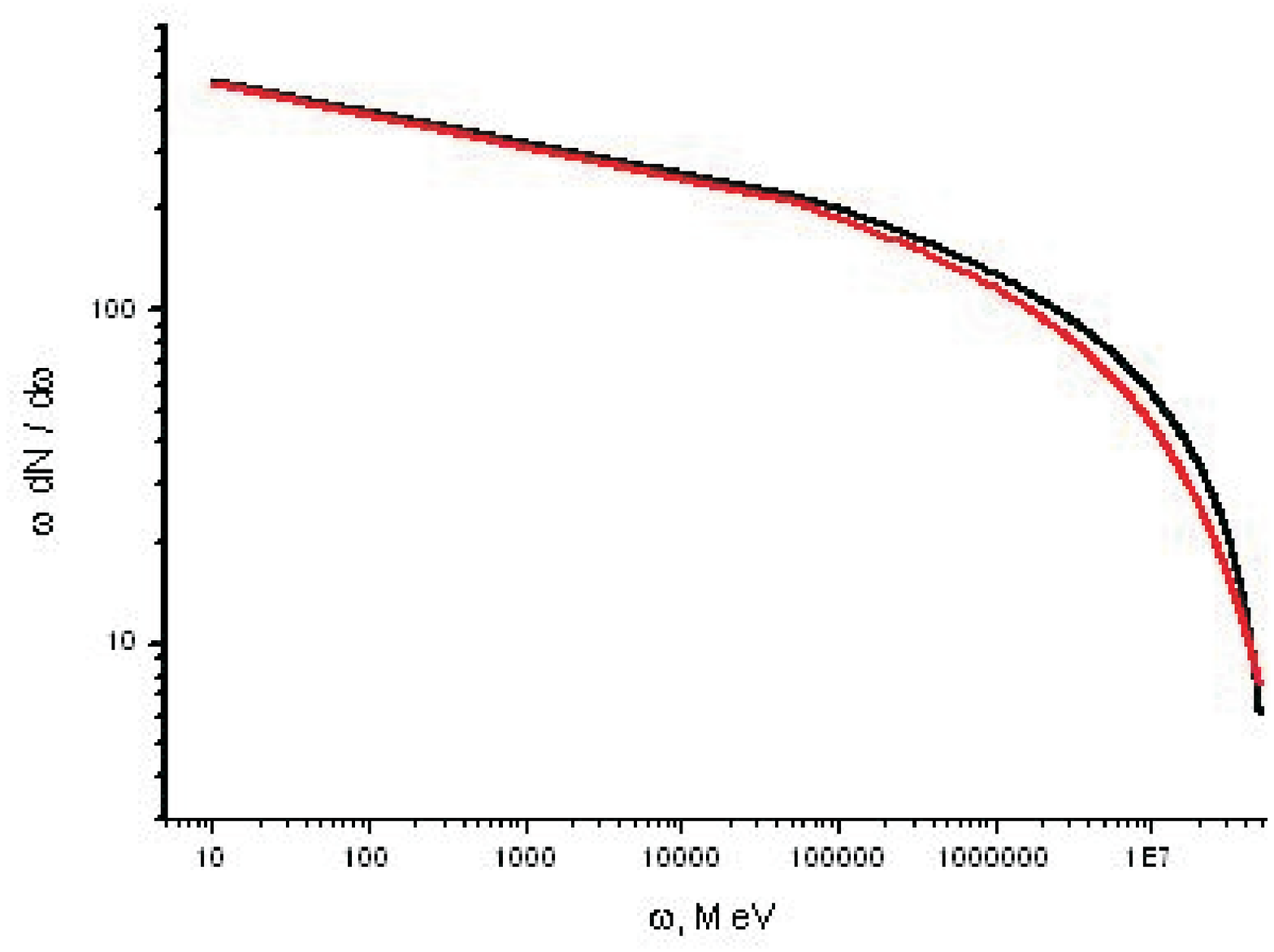}}
\caption{Спектр виртуальных фотонов по Грибову  (\ref{vf2}) ---
сплошная линия, и по методу Фурье-преобразования потенциала
Лиенара---Вихерта, проинтегрированный по прицельным параметрам $R>2R_0$, --- пунктир.} \label{f1vf}
\end{figure}

В результирующем потоке виртуальных фотонов (ВФ) можно выделить  две компоненты: продольную и перпендикулярную к оси столкновения ядер. Вследствие колоссального лоренцевского усиления, при энергиях БАКа продольная компонента оказывается много больше перпендикулярной, так что вкладом последней можно пренебречь. 
Для продольного излучения она почти постоянна в широком диапазоне энергий $\omega \lesssim \omega_b = \gamma_L /(2R_0) \approx 4.3$~ТэВ \cite{J}. Формфактор  $1/(2R_0)$ возникает из требования целостности налетающего ядра;
$\gamma_L$ отвечает лоренцевскому усилению. При больших энергиях продольная компонента экспоненциально убывает. В целях большей наглядности, на рис. \ref{f1vf} мы приводим график $\omega n(\omega)$.  Принципиально другой характер у перпендикулярной компоненты: она группируется возле значений $\omega \approx \omega_b$, хотя и вблизи максимума ее вклад остается несущественным.

\section {Оценка сечения фотопоглощения жесткого гамма-кванта ядром}

В зависимости от энергии налетающего гамма-кванта, расчет сечения взаимодействия с ядром можно выделить несколько областей по энергии, в которые различаются по механизмом взаимодейтствия и величиной сечения. При низких энергиях до двух -- трех десятков МэВ, сечение взаимодействия определяется гигантским дипольным резонансом (ГДР).
    Сечение фотопоглощения в состояние ГДР достаточно хорошо описывается в одноуровневом приближении в брэйт-вигнеровской форме, если мы сопоставим ему уровень с энергией  $\omega_0$ и шириной $\Gamma$:
\be
\sigma_{GDR} = \frac{16}{9} \alpha \pi^3 \omega B(E1; 0\to \omega_0)
    \frac{\Gamma/2\pi}{(\omega-\omega_0)^2 +(\Gamma/2)^2}\,, \label{GDR}
\ee
где $B(E1; 0\to \omega_0)$ ---  приведенная электромагнитная ширина для перехода ядра из основного состояния в состояние ГДР. В одноуровневой форме (\ref{GDR}) ее можно определить из энергетически-взвешенного правила сумм (EWSR):
\be
\omega_0 B(E1; 0\to \omega_0) = 14.8 \frac{NZ}{A} fm^2\ MeV\,.  \label{BE1}
\ee
Экспериментальные энергии ГДР хорошо описываются соотношением
$\omega_0 = 80/A^{1/3}$~МэВ. Для $^{208}$Pb $\omega_0$ =  13.5 МэВ. Экспериментальные сечения фотопоглощения на ядре  $^{208}$Pb хорошо изучены (\cite{GDR1,GDR2,GDR3} и литература, цитированная там).  Приведенные выше соотношения хорошо воспроизводят приведенные данные с шириной $\Gamma$ = 4 MeV. В пике ГДР сечение достигает 600 мбарн.
    Принимая во внимание энергию связи нейтрона порядка 7 -- 10 МэВ, понятно, что ГДР распадаются с испусканием  1 -- 3 нейтронов,  с учетом ширины ГДР. Эмиссия протона гораздо менее вероятна. Спектр испущенных нейтронов максвелловский, с температурой $\sim$1~МэВ \cite{GDR1,GDR2,GDR3,nedorez,pshen1,pshen2}.

В диапазоне энергий гамма-квантов 30 -- 150 МэВ, механизм поглощения меняется. При этих энергиях доминирует квазидейтронный механизм поглощения фотона парой нуклонов \cite{lev,nedorez}.  Сечение на ядре свинца почти постоянно, и составляет приблизительно 15 мб. В результате поглощение фотона парой нуклонов приводит к эмиссии одного-двух быстрых нуклонов, однако значительная часть энергии остается запасенной в ядре. Это делает возможной эмиссию до десятка испарительных нейтронов.

    В области энергий $\sim$300 МэВ наблюдается широкий максимум, с полушириной около 300 МэВ, соответствующий рождению в ядре $\Delta$-изобары. Сечение в пике возрастает до 88 миллибарн.

    При дальнейшем возрастании энергии фотона, вплоть до $\sim$100 МэВ, сечение стабилизируется на уровне 15 мб.
    При энергиях гамма-квантов  сотни ГэВ и выше, при которых отсутствует прямая информация о сечениях фотон-ядерного взаимодействия, это сечение мы параметризовали в форме
\be
 \sigma_{\gamma n} \sim  s^{0.1} \,.   \label{as1}
\ee
Нормировка в абсолютном сечении (\ref{as1}) определялась путем подгонки к экспериментальным данным при меньших энергиях.

\section{Результаты расчета сечений кулоновского возбуждения одного из налетающих ионов}

Несмотря на то, что кинематическая область энергий виртуальных
фотонов очень велика, как отмечено выше, взаимодействие в каждой
из подобластей обладает своими характерными особенностями и своей
величиной, поэтому вклад от разных кинематических областей
оказывается существенно различным. Это создает предпосылки для
идентификации этих механизмов взаимодействия в эксперименте.
Используя формулу (\ref{vf1}), определим интегральную величину
сечения фотоабсорбции одним из сталкивающихся ядер фотона с
энергией в диапазоне $E_a \leq \omega \leq E_b$: \be \sigma =
\int_{E_a}^{E_b}\frac {d\sigma}{d\omega} \ d\omega  \,.
\label{vf3} \ee Вычисленные вклады различных областей в полное
сечение приведены в таблице.
\begin{table}
\caption{Вклады отдельных кинематических областей энергий виртуальных фотонов $[E_a,E_b]$ в функцию возбуждения ядра-мишени}
\begin{tabular}{||c|c|c|c|c|c||}
\hline \hline
   $E_a$, MeV & $E_b$, MeV &  cross-section, b  & relative contribution, \%
& $N_n$ & $N_p$ \\
\hline
        8    &   30    &115.89  &   60.85  &  1 -- 3  &  \\
       30    &  150   &  10.50  &   5.51   &  3 -- 12  & 0.5  \\
      150    & 1000  &   35.04  &   18.39  &  2 -- 18& 0.5   \\
   1000  &  10000  &   11.05   &     5.80   &    11 -- 31 & 1 -- 3  \\
    10000   &100000 &      7.66  &  4.02      &      14 -- 38 &  1 -- 3 \\
   100000 & 1000000   &    5.68  &   2.99     &  17 -- 45  &  2 -- 4   \\
 1000000 &10000000   &     3.65  &  1.91  &  20 -- 51  &  2 -- 4 \\
10000000 &60000000 &     0.97  &  0.51    &  22 -- 55  & 3 -- 5   \\
\hline \hline
\end{tabular}
\label{pcst}
\end{table}
Из приведенных результатов следует, что доминирующий вклад в
кулоновское возбуждение происходит от ГДР:  64.6\%.
\hyphenation{Со-от-вет-ст-ву-ю-щее} Соответствующее сечение
составляет 115 барн. Наиболее вероятный канал распада --- эмиссия от одного до
трех испарительных нейтронов. Более детально, вклад области ГДР в процессы с \begin{table}
\caption{\footnotesize Парциальные сечения фотодезинтеграции одного иона с испарением от одного до четырех нейтронов в области возбуждения ГДР от 8 до 40 МэВ}
\begin{tabular}{||c||c|c|c|c|c||}
\hline  \hline
$N_n$  & 1 & 2 & 3 & 4 & $total$    \\
\hline
$\sigma_n$, barn    &   91.10 & 23.83 & 2.79 & 0.67 & 118.39    \\
\hline  \hline
\end{tabular}
\label{Pnt}
\end{table}
испарением определенного числа нейтронов представлен в таблице \ref{Pnt}. Расчет выполнен по статистической модели с параметрами \cite{BM}. Из приведенных в ней данных следует, что вплоть до энергии воэбуждения 40 МэВ вероятность испарения четырех нейтронов остается малой. Доминирующим остается канал с испарением одного нейтрона, вклад которого более 70\%. Около 20\% приходится на испарение двух нейтронов, и около 3\%  вносит трехнейтронный распад. Эти результаты хорошо согласуются с появившимися недавно экспериментальными значениями \cite{alice}.

Область квазидейтронного поглощения
дает относительно малый вклад: 5.9\%. Однако поглощение фотонов в
этой области сопровождается эмиссией в несколько раз большего
числа нейтронов при каждом акте взаимодействия. Заметный вклад в
сечение фотодезинтеграции происходит от резонансной области возбуждения
$\Delta$-изобары: около 20\%. В этом случае испарительный каскад
может сопровождаться эмиссией десятков нейтронов. Более детально
их число оценивается в следующем разделе. С дальнейшим ростом
энергии фотона дифференциальное сечение (\ref{vf1}) значительно
падает как $1/\omega$ (рис. \ref{f1vf}). Однако это падение почти
компенсируется колоссальным увеличением фазового объема, вплоть до
предельных значений энергий виртуальных фотонов $\lesssim$5~ТэВ.
Поэтому результирующий вклад этих кинематических областей в
сечение оказывается также весьма заметным: 29 барн, что составляет
$\sim$15\% от полного сечения кулоновского возбуждения.
Замечательно, что поглощение таких фотонов должно сопровождаться
эмиссией уже десятков испарительных нейтронов и до десятка быстрых
нуклонов, а также одного -- двух десятков заряженных мезонов.  В
последних двух колонках приводится оценка для числа испущенных
нейтронов и протонов, включая для простоты быстрые и испарительные
частицы. При энергиях фотонов до 1000 МэВ мы учитвали расчет
среднего числа испарительных нейтронов и их дисперсии, согласно
рис. 3 из работы \cite{pshen1}. При энергиях $E_\gamma >$ 1000 МэВ
оценки были выполнены, как описывается в следующем разделе. Важно
отметить, что из приведенных результатов следует принципиальная
\hyphenation{рез-ко-а-сим-мет-рич-ных} возможность наблюдения
$\emph{резко-асимметричных}$ событий на двух сталкивающихся ионах, в
которых наблюдается эмиссия двух-трех испарительных нейтронов
одним ядром и десятков нейтронов плюс значительное число быстрых
заряженных частиц другим ядром.  В следующем разделе произведем
более детальные оценки множественности и спектра испускаемых
частиц в случае поглощения энергичных виртуальных фотонов.

\section{Оценка множественности эмиссии нейтронов и заряженных частиц при поглощении ядром жестких гамма-квантов}

Принято считать, что при энергиях гамма-кванта порядка ГэВ и выше доминирует однонуклонный механизм взаимодействия его с ядром (например, \cite{nedorez}).
В результате взаимодействия вылетает один или два быстрых нуклона, после чего энергия оставшихся дырок диссипируется и сбрасывается в виде пред- и испарительрного каскада.
Отсюда можно предположить, что с увеличением энергии фотона количество испущенных нуклонов относительно слабо возрастает, поскольку энергия образующихся дырок остается существенно та же. Это подтверждается каскадным расчетом. Так, из результатов работы  \cite{pshen1}, рис. 3 следует, что с увеличением  энергии фотона значительно падает доля энергии, остающейся в ядре после выбивания быстрого нуклона.  В итоге, при взаимодействии с ядром фотона с энергией 10 ГэВ предсказывается эмиссия в среднем только порядка двадцати нейтронов, хотя энергия фотона более чем достаточна для поляной дезинтеграции ядра.  Эти оценки даже еще более усугубляются в работах \cite{mar1, mar2}, в которых предсказывается испускание не более нескольких нуклонов при взаимодействии ядра с фотоном даже при  энергии порядка 200 ГэВ. Вероятно, вопрос динамики взаимодействия жестких фотонов с ядрами может получить дальнейшее освещение на БАКе.

    Рассмотрим качественно возможную картину взаимодействия фотна с ядром в рамках подхода Глаубера---АГК---Бертокки и Трелиани \cite{AGK, BT}. Прежде заметим, что закон сохранения энергии-импульса делает невозможным поглощение жесткого фотона одним нуклоном. Так, при энергии фотона порядка 10 ГэВ с необходимостью происходит рождение двух мезонов.  Более того, при таких энергиях взаимодействие фотона с нуклонами происходит через посредство кварк-антикварковой пары. В явом виде эта картина эксплуатируется в модели векторной доминантности, предполагающей превращение фотона в $\rho$-мезон.  Поэтому сечение взаимодействия фотона с отдельными нуклонами в ядре нельзя считать малым.  Необходщимо учитывать процессы экранировки в результате множественных взаимодействий кварк-антикварковой пары с несколькими нуклонами одновременно.  Для качественного рассмотрения заменим фотон на пи-мезон (с константой связи $\alpha \pi$). Тогда полное сечение неупругого взаимодействия можно представить в терминах посдедовательных столкновений с нуклонами ядра в виде ряда по числу ``раненых'' нуклонов \cite{BT}:
\bea
\sigma = \sum_{n=1}^{A} \sigma_n\,;
\nonumber \\
\sigma_n = \frac{A!}{(A-n)!\ n!}
    \int d^2b\ [\sigma_{in} T(b)]^n [1-\sigma_{in} T(b)]^{A-n}\,,   \label{ber}
\eea
Здесь  $\sigma_{in}$ --- полное сечение отдельного пион-нуклонного взаимодействия.  Для $E_\gamma$ = 1 ГэВ
оно составляет 20 мб, а для $E_\gamma$ = 1 ТэВ --- 40 мб \cite{pdg}. Далее,  $T(b)$ --- профильная функция прицельного параметра $b$:
\be
T(b) = \int \rho(z) \ dz\,.
\ee
Каждое из слагаемых в разложении (\ref{ber}) можно рассматривать как соответствующее физическому процессу, в котором происходит взаимодействие адрона с $n$ нуклонами, а остальные $A-n$ нуклонов остаются незатронутыми.
   Среднее число выбитых быстрых нуклонов можно найти с помощью правила сумм \cite{BT}:
\be
\bar n = \frac{\sum_n n\sigma_n}{\sum_n \sigma_n } =
\frac{A\sigma_{in}}{\sigma_{tot}}\,.    \label{SR}
\ee
 Результаты расчета по формуле (\ref{SR}) приведены в Таблице \ref{pit}. Согласно Таблице, получим, что $\bar n$ = 3.08 и 5.33 для представительных энергий фотонов  1 ГэВ и 1 ТэВ, соответственно, из них 1.2 и 2.1 протонов, считая их число пропорционально $Z/A$. Каждый выбитый нуклон оставляет ``дырку'',  энергия которой в зависимости от номера оболочки может быть порядка  $\sim$40 МэВ. После диссипации энергии дырочных состояний дальнейший распад ядра проходит испарительным каскадом, от одного до $\sim$четырех
\begin{table}
    \caption{Моделирование сечения взаимодействия гамма кванта с ядром $^{208}$Pb в зависимости от числа взаимодействий с отдельными нуклонами $n$}
\begin{tabular}{||c||c|c||c|c||}
\hline \hline
$n$ &  \multicolumn{2}{c||}{$E_\gamma$ = 1 ГэВ} &  \multicolumn{2}
{c||}{$E_\gamma$ = 1 ТэВ}   \\
\hline
&$\sigma_n$, мб&$\sigma$, мб&$\sigma_n$, мб&$\sigma$, мб    \\
1  &          0.869D+01  &       0.869D+01&          0.706D+01  &       0.706D+01   \\
    2  &          0.615D+01  &       0.148D+02&          0.410D+01  &       0.112D+02   \\
    3  &          0.516D+01  &       0.200D+02&          0.337D+01  &       0.145D+02   \\
    4  &          0.411D+01  &       0.241D+02&          0.314D+01  &       0.177D+02   \\
    5  &          0.295D+01  &       0.271D+02&          0.307D+01  &       0.207D+02  \\
    6  &          0.189D+01  &       0.289D+02&          0.299D+01  &       0.237D+02  \\
    7  &          0.107D+01  &       0.300D+02&          0.267D+01  &       0.264D+02  \\
    8  &          0.558D+00  &       0.306D+02&          0.238D+01  &       0.288D+02  \\
    9  &          0.263D+00  &       0.308D+02&          0.204D+01  &       0.308D+02  \\
   10  &          0.113D+00  &       0.309D+02&          0.165D+01  &       0.325D+02  \\
   11  &           0.443D-01  &       0.310D+02&          0.125D+01  &       0.337D+02  \\
   12  &           0.160D-01  &       0.310D+02&          0.875D+00  &       0.346D+02  \\
   13  &           0.538D-02  &       0.310D+02&          0.573D+00  &       0.352D+02  \\
   14  &           0.168D-02  &       0.310D+02&          0.352D+00  &       0.355D+02  \\
   15  &           0.491D-03  &       0.310D+02&          0.203D+00  &       0.357D+02  \\
   16  &           0.134D-03  &       0.310D+02&          0.110D+00  &       0.358D+02  \\
   17  &           0.347D-04  &       0.310D+02&           0.566D-01  &       0.359D+02  \\
   \hline  \hline
\end{tabular}
\label{pit}
\end{table}
нейтронов на ``дырку''. Пусть в среднем испаряется два нейтрона на
``дырку''. Тогда получим, оценивая дисперсию по $n$ из Таблицы
\ref{pit}, что будет вероятным испускание порядка от 6 до 26 и от
11 до 42 испарительных, и до порядка 8 и 13 быстрых нейтронов и
протонов при данных энергиях кулоновского фотона, соответственно.
Как сказано выше, учитывая результаты таблицы \ref{pcst},
вероятность таких событий можно оценить в $\sim$20\%.

\section{Заключение}

Основной вывод настоящей работы состоит в том, что

1) Приблизительно в 20\% случаев, происходит возбуждение одного из сталкивающихся ядер жестким виртуальным фотоном с энергией $\omega \gtrsim$ 1 ГэВ. Поглощение фотона сопровождается образованием в среднем трех -- пяти дырок в нуклонных оболочках, создаваемых на месте выбитых быстрых нуклонов. Вероятным каналом распада  возбужденного таким образом ядра оказывается эмиссия нескольких десятков испарительных нейтронов. В лабораторной системе координат испарительные нейтроны попадают в счетчики нулевого угла (ZDC). Выбитые быстрые нуклоны имеют промежуточное рапидити, и потому могут быть легко отличимы в эксперименте. Более того, выбивание каждого быстрого нуклона должно сопровождаться эмиссией нескольких мезонов с промежуточным рапидити, общее число которых таким образом может быть более десяти.

2) Оценка запасенной в ядре энергии является скорее $\emph{нижней границей}$. Не учитывались такие предравновесные процессы, когда выбитые фотоном быстрые нуклоны могут зацепить соседние нуклоны, передав им часть энергии, которая будет затем диссипирована ядром. Можно ожидать, что такие процессы будут весьма существенны \cite{mar1}. Более того, представляется вполне вероятным  образование в ядре нуклонного резонанса, при дальнейшем распаде которого высвободится несколько сотен МэВ, что в свою очередь может привести к испусканию десятка и более испарительных нейтронов.

3) Следует отметить неизбежную асимметрию описанного процесса на обоих ядрах. Из общих физических соображений можно ожидать появления таких событий, когда на одном ядре будут испущены по максимуму порядка сорока испарительных нейтронов с близким к нулю рапидити, и порядка десяти и более нуклонов и мезонов с промежуточным рапидити. В это время на ядре-партнере возможны события с испусканием одного -- трех испарительных нейтронов, например, от ГДР, и двух -- трех заряжнных частиц с промежуточным рапидити. Испускание последних возможно как вследствие взаимодействия жестких фотонов с ядром, так и в результате дифракционного взаимодействия отдельных нуклонов. Последний процесс мы пока не рассматриваем в деталях.

4) Указанная асимметрия может оказаться даже более сильной при учете возможных процессов поглощения двух виртуальных нуклонов одним ядром \cite{pshen1,pshen2,br}. До сих пор в экспериментах на ускорителях предыдущего поколения не найдено определенных указаний на проявление процессов двукратного взаимодействия. Считается, что их относительная вероятность составляет несколько процентов. Вклад их однако растет с энергией. Поэтому представляет интерес рассмотреть эти процессы более подробно в исследованиях на БАКе.

5)  Проверка указанных соотношений на БАКе представляет большой интерес  как тест нашего понимания  динамики взаимодействия жестких гамма-квантов с ядром.

\end{document}